%% file: main.tex
\documentclass[sigconf]{acmart}

\AtBeginDocument{%
  } 

\acmPrice{15.00}
\acmISBN{978-1-4503-XXXX-X/18/06}
\usepackage{enumitem}

\usepackage{amsmath,amssymb,amsfonts}
\usepackage{amsthm}

\usepackage{xcolor}
\usepackage{footnote}
\usepackage{multirow}  
\usepackage{algorithm}
\usepackage{algpseudocode}
\usepackage{booktabs}
\usepackage{colortbl}
\usepackage{multirow}
\usepackage{footnote}
\usepackage{tablefootnote}
\usepackage{float}
\usepackage{threeparttable}
\usepackage{pifont}
\usepackage{textcomp}
\usepackage{tabularx}
\usepackage{graphicx}

\hypersetup{
  colorlinks,
  citecolor=violet,
  linkcolor=red,
  urlcolor=blue}

\makeatletter
\let\@ACM@copyright@check@cc\relax
\makeatother
\setcopyright{cc}
\setcctype{by}

\copyrightyear{2026}
\acmYear{2026}
\acmConference[DAC '26]{63rd ACM/IEEE Design Automation Conference}{July 26--29, 2026}{Long Beach, CA, USA}
\acmBooktitle{63rd ACM/IEEE Design Automation Conference (DAC '26), July 26--29, 2026, Long Beach, CA, USA}
\acmDOI{10.1145/3770743.3804042}
\acmISBN{979-8-4007-2254-7/2026/07}

\begin{document}


\title{DRIFT: Harnessing \underline Inherent \underline Fault \underline Tolerance for Efficient and \underline Reliable \underline Diffusion Model Inference} 

\author{Jinqi Wen$^{13}$\footnotemark[1], Tong Xie$^{21}$\footnotemark[1], Runsheng Wang$^{245}$, and Meng Li$^{124}$\footnotemark[2]}


\affiliation{%
  \institution{\textit{$^1$Institute for Artificial Intelligence \& $^2$School of Integrated Circuits, Peking University, Beijing, China}}
  \country{}
}
\affiliation{%
  \institution{\textit{$^3$School of Electronics Engineering and Computer Science, Peking University, Beijing, China}}
  \country{}
}
\affiliation{%
  \institution{\textit{$^4$Beijing Advanced Innovation Center for Integrated Circuits, Beijing, China}}
  \country{}
}
\affiliation{%
  \institution{\textit{$^5$Institute of Electronic Design Automation, Peking University, Wuxi, China}}
  \country{}
}

\newcommand{\ml}[1]{{\color{red}\bf [Meng: #1]}}

\newcommand{\xt}[1]{{\color{orange}\bf [xt: #1]}}
\newcommand{\wjq}[1]{{\color{blue}\bf [wjq: #1]}}


\setlength{\headsep}{10pt} 

\setlength{\footskip}{17pt} 

\addtolength{\textheight}{0.25in} 

\addtolength{\voffset}{-0.15in}

\begin{abstract}
Diffusion model deployment has been suffering from high energy consumption and inference latency despite its superior performance in visual generation tasks. Dynamic voltage and frequency scaling (DVFS) offers a promising solution to exploit the potential of the underlying accelerators. However, existing approaches often lead to either limited efficiency gains or degraded output quality because they overlook the inherent fault tolerance of the diffusion model. 
Therefore, in this paper, we propose DRIFT, a novel algorithm-architecture co-optimization framework that harnesses the fault tolerance for efficient and reliable diffusion model inference. We first perform a comprehensive resilience analysis on representative diffusion models. 
Building on these observations, we introduce a fine-grained, resilience-aware DVFS strategy that selectively protects error-sensitive network blocks and timesteps, and a rollback algorithm-based fault tolerance (ABFT) mechanism that adaptively corrects only critical errors by reverting to previous timesteps. We further optimize offloading intervals and reorganize data layouts to reduce memory overhead. Experiments across diverse models and datasets show that DRIFT can achieve on average 36\% energy savings through voltage underscaling or 1.7$\times$ speedup via overclocking while maintaining generation quality.

\vspace{-3pt}
\vspace{2ex} 
\noindent\textbf{ACM Reference Format:}\par\noindent
Jinqi Wen, Tong Xie, Runsheng Wang, and Meng Li. 2026. DRIFT: Harnessing Inherent Fault Tolerance for Efficient and Reliable Diffusion Model Inference. In \textit{63rd ACM/IEEE Design Automation Conference (DAC '26), July 26--29, 2026, Long Beach, CA, USA}. ACM, New York, NY, USA, 7 pages. \url{https://doi.org/10.1145/3770743.3804042}
\end{abstract}



\maketitle

\begingroup
\renewcommand{\thefootnote}{\fnsymbol{footnote}}

\makeatletter
\makeatother
\footnotetext{This work was supported in part by the National Natural Science Foundation of China (NSFC) under Grant 62125401, 62495102, 92464104, and 62341407; in part by the National Key Research and Development Program under Grant 2024YFB4505004; in part by the Beijing Municipal Science and Technology Program under Grant Z241100004224015; in part by the Beijing Outstanding Young Scientist Program under Grant JWZQ20240101004; and in part by the 111 Project under Grant B18001.}
\footnotetext[1]{Equal contribution. \hspace{1.5em} $^\dagger$Corresponding author: meng.li@pku.edu.cn}
\endgroup

\input{docs/1_Introduction}

\input{docs/2_Background}

\input{docs/3_Framework}

\input{docs/4_Resilience}

\input{docs/5_Method}
\input{docs/6_Evaluation}

\input{docs/7_Conclusions}
\newpage
\bibliographystyle{ieeetr}
\bibliography{top_simplified.bib,reference_compressed}

\end{document}

%% file: docs/1_Introduction.tex
\vspace{-9pt}
\section{Introduction}
\vspace{-2pt}
Diffusion models have emerged as a leading class of generative models, achieving remarkable performance in various fields such as image generation \cite{ho2020denoising}. Featuring a backward inference process that iteratively denoises a Gaussian sample, diffusion models typically require tens to hundreds of iterations (i.e., \textit{timesteps})
per generation \cite{songdenoising}.
While enabling superior performance, 
such a process incurs significant energy consumption \cite{ jing2024aig} and inference latency \cite{kim2025ditto}, hindering the widespread deployment of diffusion model.

\begin{figure}[!tb]
    \centering
    \includegraphics[width=0.9\linewidth]{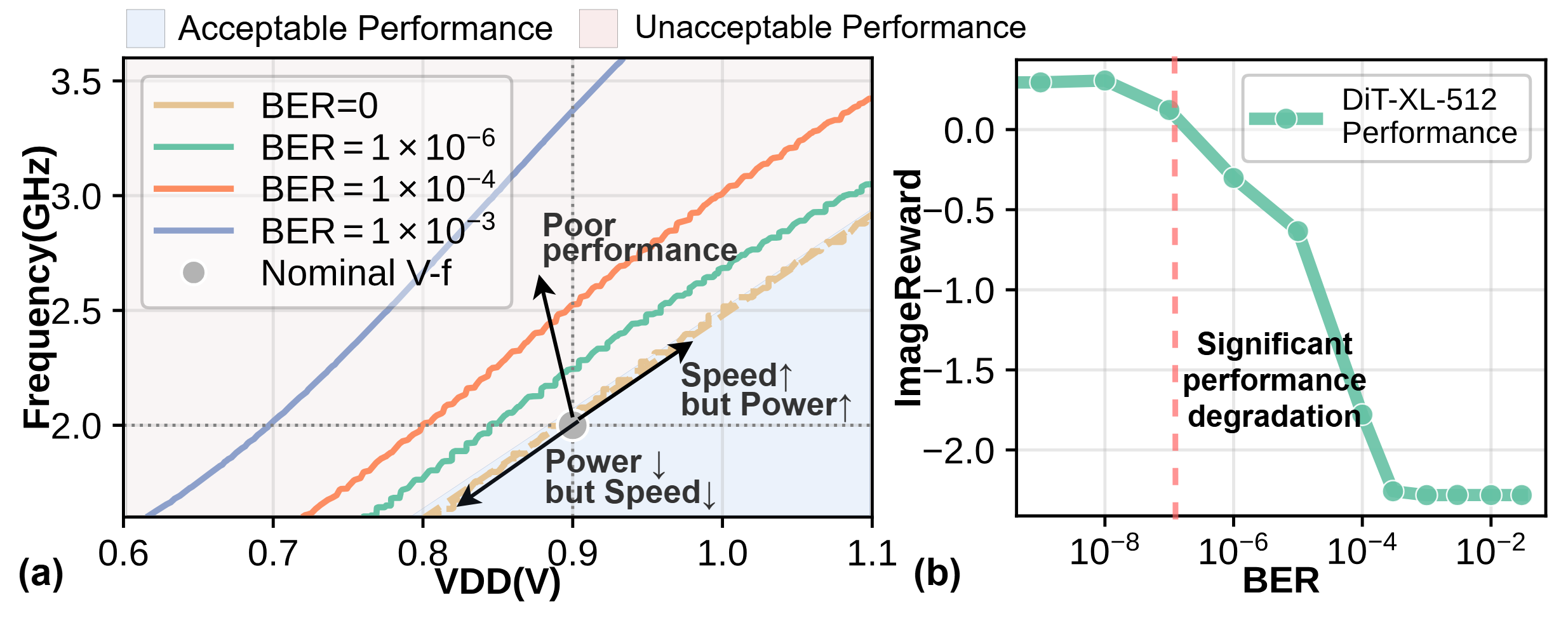}
    \vspace{6pt}
    \caption{
    Limitations of applying DVFS to diffusion generation.
    (a) BER on different operating points (synthesized with commercial 14nm PDK). (b) Increased BER leads to significant model performance degradation. 
    }
    \vspace{-10pt}
    \label{fig:1challenge}
\end{figure}

While many customized accelerators \cite{kong2024cambricon, qi2025mhdiff, park2025radit, kim2025ditto, jing2024aig, heo2025exion} have been proposed to improve diffusion inference through designing specific dataflow and reducing redundant computation, dynamic voltage and frequency scaling (DVFS) remains underexplored. As diffusion inference is typically compute-intensive \cite{kim2025ditto, park2025radit, du2025fewer}, this is a promising direction to reduce inference energy and latency.
However, conventional DVFS is tightly constrained by timing requirements, limiting achievable efficiency gains. As in Fig.~\ref{fig:1challenge}(a), improved energy efficiency often has to come at the cost of reduced throughput.
More aggressive DVFS can induce timing violations and increase computation bit error rates (BERs), ultimately degrading generation quality (Fig.~\ref{fig:1challenge}(b)).
Although many approaches have been proposed to mitigate timing errors \cite{khoshavi2020shieldenn, libano2018selective, ernst2003razor, zhang2018thundervolt, gundi2020effort, pandey2019greentpu, zhang2019fault, hsiao2023mavfi}, such as algorithm-based fault tolerance (ABFT) \cite{huang1984algorithm, xue2023approxabft, bal2023novel, xie2025realm}, none of them can balance efficiency and reliability well in the context of diffusion model inference.

Nonetheless, similar to other deep neural networks (DNNs), diffusion models are expected to exhibit inherent error resilience. For example, in Fig.~\ref{fig:1challenge}(b), the generation quality does not degrade at low BERs, suggesting ample opportunities to improve efficiency through aggressive DVFS without sacrificing generation quality. 
However, the resilience characteristics of diffusion models are not well understood, hindering their strategic exploitation.
Considering the unique multi-step denoising process and image-based outputs, the resilience behavior of diffusion generation could differ significantly from other DNNs \cite{reagen2018ares, li2017understanding, mahmoud2021optimizing, wan2021analyzing, wan2022frl, agarwal2023resilience, xie2025realm}.

To this end, we propose DRIFT, an algorithm-architecture co-design framework that addresses a fundamental question: \textit{can we harness the inherent resilience of diffusion models to achieve efficient and reliable inference?} To answer this, we first conduct a large-scale error injection study to comprehensively characterize the resilience behavior of diffusion models. Our analysis reveals that errors in lower bits have a marginal impact, and that robustness varies across network blocks and denoising timesteps. Moreover, we observe self-correction behavior, where later steps partially compensate for earlier errors. Building on these findings, we introduce (i) a fine-grained DVFS strategy that performs module- and timestep-specific voltage-frequency scaling based on heterogeneous resilience, and (ii) a rollback-ABFT mechanism that filters large errors via ABFT and recovers them with corresponding values from previous timesteps. We further apply checkpointing at intervals and data repacking to minimize memory access overhead. In this way, DRIFT effectively mitigates error impacts during inference, allowing for more aggressive DVFS without sacrificing generation quality. Our contributions can be summarized as follows:

\vspace{-3pt}
{\setlength{\leftmargini}{1em}
\begin{itemize}
    \item We perform an error injection study on representative diffusion models to analyze the impact of DVFS-induced timing errors. Our results reveal heterogeneous robustness across network blocks and denoising timesteps, as well as a self-correcting behavior.
    \item Based on our characterization, we propose DRIFT, an algorithm-architecture co-design framework that incorporates fine-grained resilience-aware DVFS and rollback-ABFT to mitigate error impacts with minimal overhead.
    \item Extensive experiments demonstrate that by enabling aggressive DVFS, DRIFT can achieve an average 36\% energy reduction through voltage underscaling or $1.7\times$ speedup via overclocking, all while preserving image generation quality.
\end{itemize}
}

%% file: docs/2_Background.tex
\vspace{-1pt}
\section{Background}

\begin{figure}[!tb]
    \centering
    \includegraphics[width=1\linewidth]{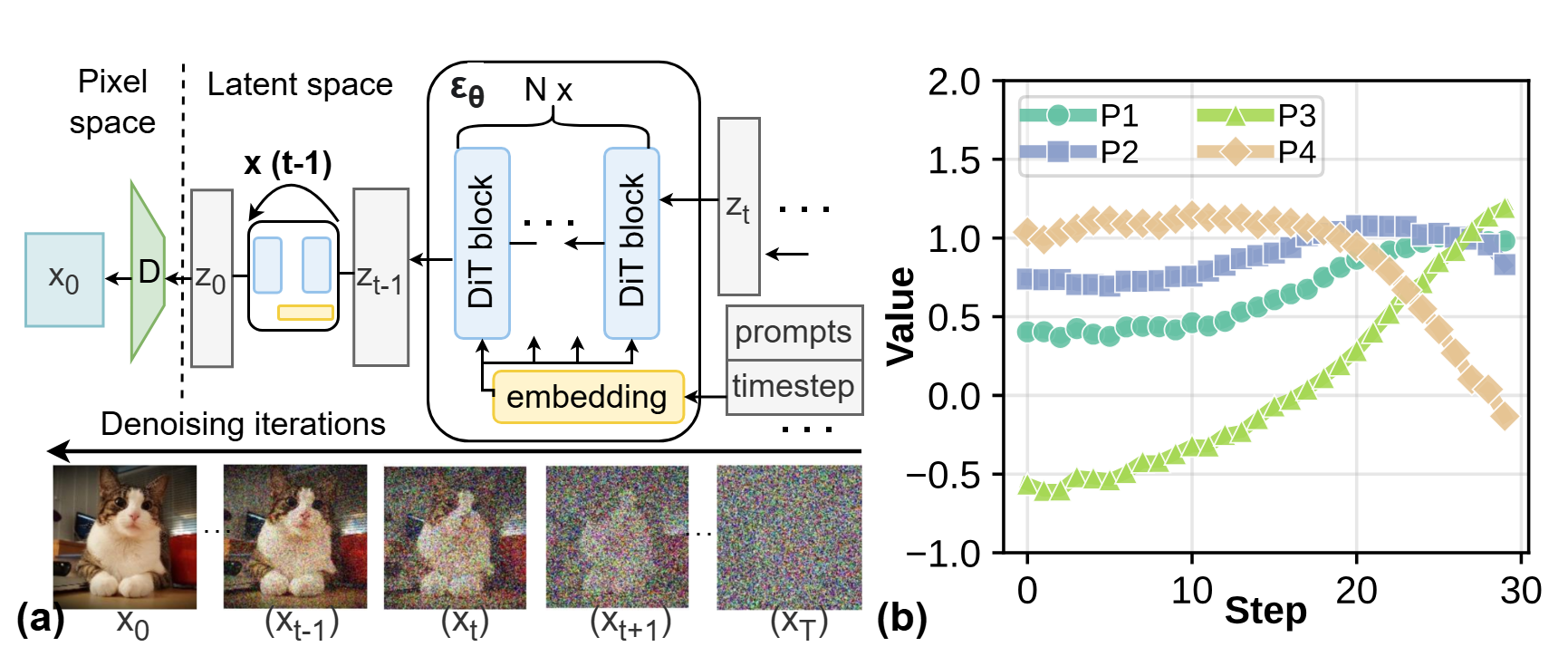}
    \vspace{-0pt}
    \caption{(a) Diffusion generation workload. (b) Similarity of activation values across adjacent steps.}
    \vspace{-3pt}
    \label{fig:2diffusion}
\end{figure}

\vspace{-1pt}
\subsection{Diffusion Model Overview}
\vspace{-1pt}
\label{bg_diffusion}
Diffusion models generate images by iteratively denoising a Gaussian sample over discrete \textit{timesteps} (or \textit{steps}) \cite{ho2020denoising}. At each step, a denoising model refines the noisy latent representation from the previous step, conditioned on a text prompt and timestep embedding, as depicted in Fig.~\ref{fig:2diffusion}(a). Popular backbones for this model include the UNet-based \cite{rombach2022high} and Transformer-based \cite{peebles2023scalable, chen2023pixart, li2024cogact} (Diffusion Transformer, or DiT) architectures.
As a compute-bound workload \cite{kim2025ditto, park2025radit, du2025fewer}, diffusion models are commonly deployed on customized accelerators such as TPU-like systolic arrays \cite{jouppi2017datacenter, qi2025mhdiff, kim2025ditto}.

Despite their impressive capacity, iterative denoising models suffer from high energy consumption and latency. Fortunately, the progressive denoising process presents an opportunity: as shown in Fig.~\ref{fig:2diffusion}(b), activations across timesteps are numerically similar. This temporal similarity has been exploited in many algorithms and accelerators. For example, 
\cite{heo2025exion, park2025radit}
supported model compression via quantization and pruning, while \cite{lu2022dpm, salimansprogressive} reduced the number of denoising steps through efficient sampling. \cite{ma2024deepcache, TaylorSeer2025, kahatapitiya2025adaptive} cached similar activations across steps, and \cite{kong2024cambricon, kim2025ditto, qi2025mhdiff} customized accelerators to compute narrow-range deltas instead of full activations. While successful in reducing computation, these methods did not explore directly switching operating points. DVFS, in combination with these existing techniques, could further boost inference efficiency.

\vspace{-2pt}
\subsection{Fault-Tolerant DNN Inference}
\vspace{-1pt}
\label{bg_fault_tolerant_nn}
\vspace{-1pt}
\paragraph{\textbf{Error Sources}}
Hardware accelerators are susceptible to permanent and transient faults caused by fabrication defects, aging, variation, and radiation-induced single-event upsets \cite{dixit2021silent, huang2017variability, moghaddasi2023dependable, jiao2017clim}. 
This paper focuses on transient computational errors arising from aggressive DVFS during inference. 
Lowering the supply voltage increases datapath delay, while raising the clock frequency shortens the clock period, both leading to timing violations and, consequently, incorrect computation results \cite{jiao2017clim, zhang2018thundervolt, bal2023novel, zhang2023read}.
\vspace{-1pt}
\paragraph{\textbf{Model Resilience Characterization}}
Model error resilience is commonly quantified by the relationship between BER and model performance \cite{reagen2018ares, wan2021analyzing}, with random bit-flip fault injection serving as a widely adopted abstraction \cite{zhang2023read, kim2019dris, sangchoolie2017one}. Prior works have examined various factors influencing CNN resilience, including quantization \cite{reagen2018ares}, data reuse \cite{li2017understanding}, and error location \cite{mahmoud2021optimizing}. Others have extended resilience analysis to reinforcement learning \cite{wan2021analyzing, wan2022frl} and large language models \cite{agarwal2023resilience, xie2025realm}. However, existing analysis of diffusion models is scarce. While \cite{gao2024dependability} compared the error resilience of different blocks, it focused solely on Stable Diffusion \cite{rombach2022high} and overlooked the inter-step behaviors. Given their iterative denoising process and the direct mapping of all output elements to the final image, diffusion models may exhibit distinct resilience behaviors.

\vspace{-1pt}
\subsection{Tradeoff between Efficiency and Reliability}
\vspace{-1pt}
\label{bg_tradeoff}
The efficiency gains from traditional DVFS techniques are often limited, as adjustments must remain within strict timing constraints \cite{cherupalli2016exploiting, le2010dynamic, zhang2023avatar}. In the presence of timing errors, although various techniques have been proposed, most are impractical for diffusion model inference. For instance, DMR \cite{khoshavi2020shieldenn, libano2018selective} is widely used for error detection but doubles the computation cost. Timing-borrowing approaches \cite{ernst2003razor, zhang2018thundervolt, gundi2020effort} employ shadow flip-flops to capture late signals but lack scalability in large-scale accelerators. Timing error prediction methods \cite{pandey2019greentpu}, which compare input patterns with calibrated suspects, are not portable across devices because they depend on hardware specifications. Anomaly detection schemes \cite{zhang2019fault, hsiao2023mavfi, xue2023approxabft} monitor data distributions to prune abnormal outliers, but lead to substantial information loss and degraded generation quality.

\begin{figure}[!tb]
    \centering
    \includegraphics[width=0.9\linewidth]{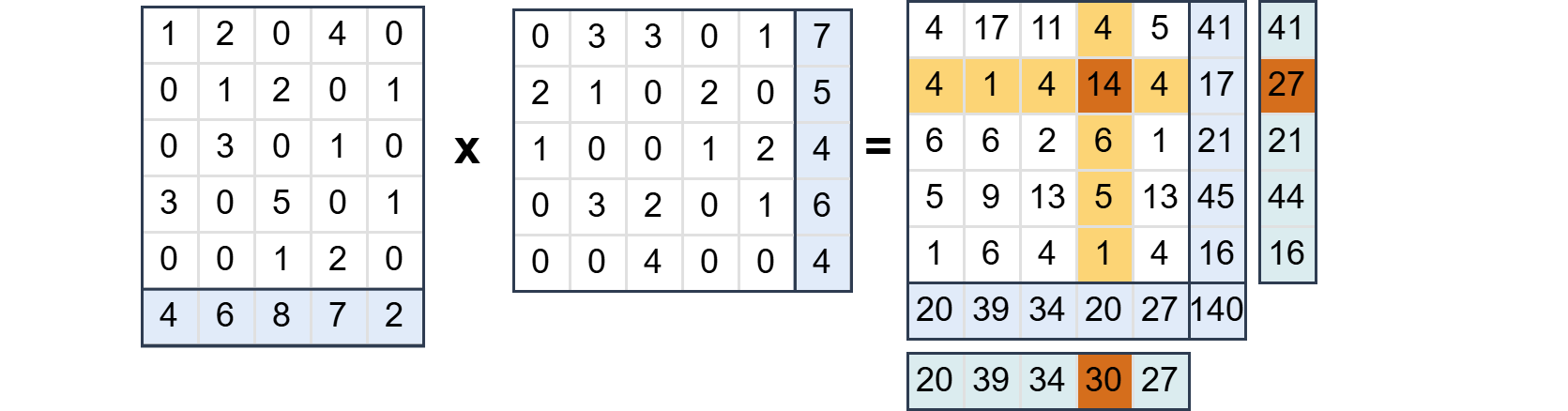}
    \vspace{12pt}
    \caption{ABFT can indicate error magnitude and location.}
    \vspace{-5pt}
    \label{fig:3ABFT}
\end{figure}
ABFT \cite{huang1984algorithm} is suitable for general matrix multiplication (GEMM), the core operation in DNNs, and can be efficiently integrated into systolic arrays \cite{bal2023novel}. As Fig.~\ref{fig:3ABFT} illustrates, by appending checksums to the operands, ABFT can help detect and locate errors.
Error recovery, e.g., recomputation, will be triggered upon error detection.
Although \cite{xue2023approxabft, xie2025realm} have explored allowing minor computational errors, this relaxation is marginal for diffusion models. Under aggressive DVFS, they encounter frequent recoveries that can erase efficiency benefits.

%% file: docs/3_Framework.tex
\section{Timing Error Modeling}
\vspace{-1pt}

\vspace{-2pt}
\subsection{Error Model}
\vspace{-2pt}
\label{frame_err_model}
Since we focus on transient computational errors during inference, we assume error-free data retrieval from memory. We model these timing errors with the widely-adopted uniform random bit-flip model \cite{sangchoolie2017one, zhang2023read, wan2021analyzing}, parameterized by the BER. 
BERs under different voltage-frequency operating points are simulated by performing timing analysis using Synopsys PrimeTime and HSPICE, incorporating toggle rates measured from real diffusion model inference. The error patterns are consistent with previously reported measurements \cite{ernst2003razor,jiao2017clim, xie2025realm}.

\vspace{-4pt}
\subsection{Error Injection Method}
\label{frame_err_inject}
\vspace{-3pt}
Our error injection framework leverages the hook mechanism in PyTorch. Following previous practice \cite{xiao2023smoothquant,kong2024cambricon, qi2025mhdiff}, the weight and input activation are quantized to INT8, and the hook function introduces bit flips into the corresponding INT32 output tensor, which is then propagated through subsequent computations. 
Each bit-flip location is identified by timestep, block, tensor index, and bit position, which can be randomly selected or explicitly specified according to requirements.

%% file: docs/4_Resilience.tex
\begin{figure}[!tb]
    \centering
    \includegraphics[width=0.9\linewidth]{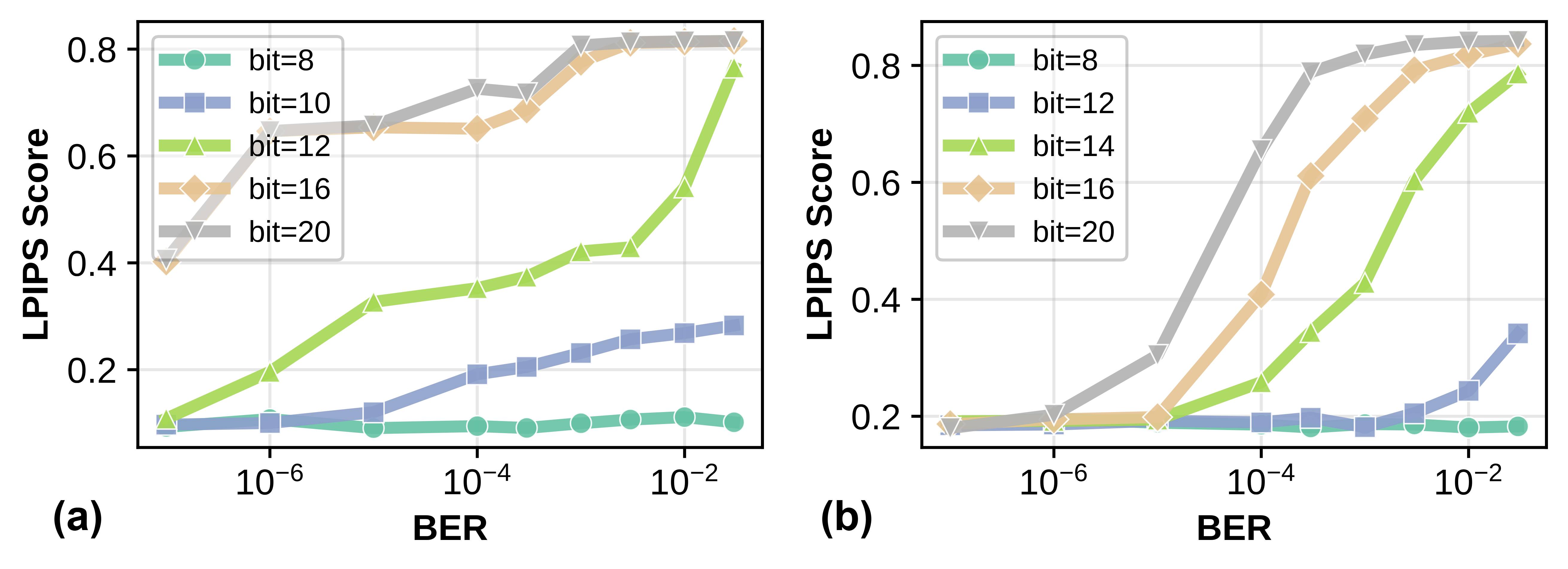}
    \vspace{7pt}
    \caption{
        {Bit-level} resilience on (a) DiT and (b) PixArt.
    }
    \vspace{-1pt}
    \label{fig: 4resilience_characterization1}
\end{figure}

\section{Resilience Characterization}
\label{fault_resilience}


This section presents our resilience characterization on diffusion models, aiming to answer the following questions:
{\setlength{\leftmargini}{1em}
\begin{itemize}
    \item \textbf{Q1}: How does the bit-flip position affect the generation quality?
    \item \textbf{Q2}: How do error impacts differ across denoising steps?
    \item \textbf{Q3}: How does error resilience vary among network blocks?
    \item \textbf{Q4}: Does the diffusion process exhibit self-correction behavior?
\end{itemize}
}

Following \cite{ma2024deepcache, kim2025ditto, jing2024aig, TaylorSeer2025}, we benchmark generation quality on several representative diffusion models.
Due to page limitations, we only present results for DiT-XL512 \cite{peebles2023scalable} on ImageNet \cite{deng2009imagenet} and PixArt-alpha \cite{chen2023pixart} on COCO val2017 \cite{lin2014microsoft}. 
More evaluations will be presented in Sec.~\ref{sec: experiments}.
Generation quality is evaluated using standard metrics, including CLIP ($\uparrow$) \cite{hessel2021clipscore}, ImageReward ($\uparrow$) \cite{xu2023imagereward}, and LPIPS ($\downarrow$) \cite{zhang2018unreasonable} score. We argue that resilience characterization should avoid comparing fundamentally different images and exclude variations between samples of the same task. Therefore, in this section, we prioritize LPIPS, which measures perceptual similarity between image patches, with fixed initial noise seeds. CLIP and ImageReward results are also reported in Sec.~\ref{experiment_quality_efficiency} for a comprehensive assessment.

\vspace{-5pt}
\subsection{Bit-Level Resilience}
\label{fault_resilience_bit}

Fig.~\ref{fig: 4resilience_characterization1} shows the resilience behavior when injecting errors to different bit positions.
We can observe that
\textbf{diffusion models tolerate small errors, and performance loss mainly arises from high-bit flips that significantly distort value magnitudes}.

\vspace{-3pt}
\subsection{Timestep-Level Resilience}
\vspace{-3pt}
\label{fault_resilience_t}

\begin{figure}[!tb]
    \centering
    \includegraphics[width=0.9\linewidth]{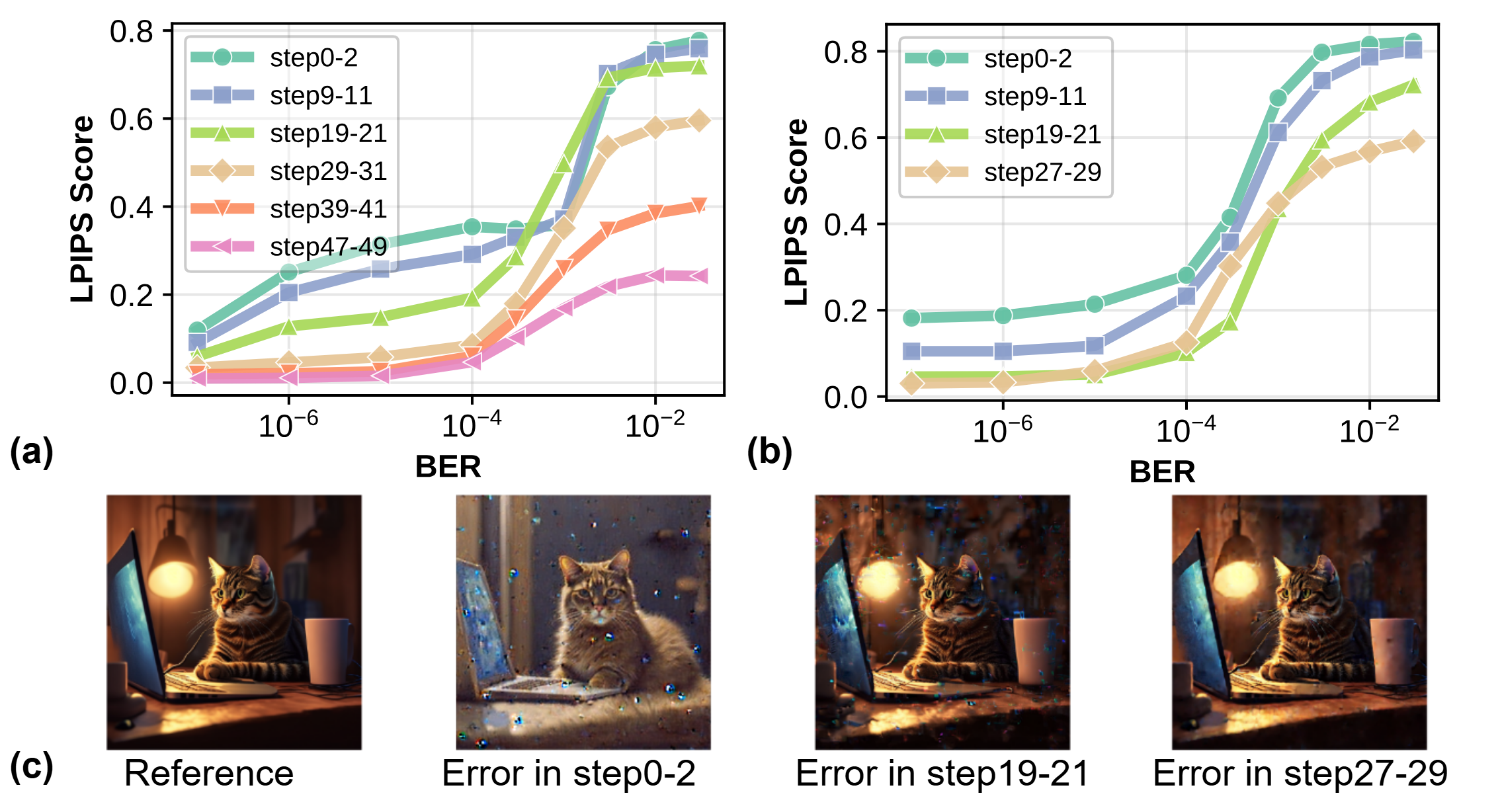}
    \vspace{9pt}
    \caption{
        Timestep-level resilience on (a) DiT and (b) PixArt. (c) Visualization example.
    }
    \vspace{2pt}
    \label{fig: 4resilience_characterization2}
\end{figure}

To analyze the temporal resilience during denoising, we inject errors at different denoising timesteps and measure the corresponding LPIPS changes in Fig.~\ref{fig: 4resilience_characterization2}. Our results indicate that \textbf{earlier timesteps are substantially more sensitive}, exhibiting significantly larger performance degradations. 
This behavior can be explained by the generative mechanism of diffusion models. Early steps are responsible for constructing global semantics and spatial structure, whereas later steps primarily refine textures and details \cite{du2025fewer}. Therefore, faults in earlier timesteps cause semantic or structural distortions, which are especially reflected in LPIPS, whereas faults in later steps typically manifest as localized noise or texture artifacts.

\vspace{-2pt}
\subsection{Block-Level Resilience}
\label{fault_resilience_block}

\begin{figure}[!tb]
    \centering
    \includegraphics[width=0.9\linewidth]{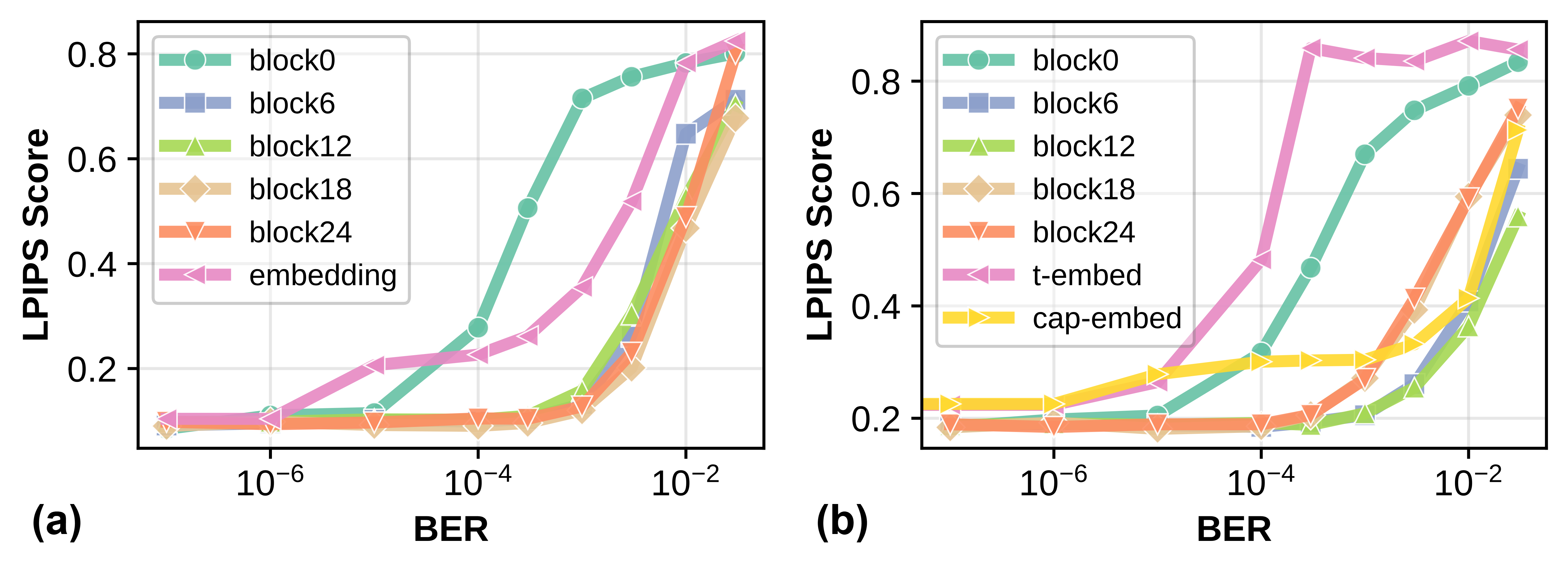}
    \vspace{5pt}
    \caption{
        {Block-level} resilience on (a) DiT and (b)PixArt}
    \vspace{-0pt}
    \label{fig: 4resilience_characterization3}
\end{figure}
\begin{figure}[!tb]
    \centering
    \includegraphics[width=0.9\linewidth]{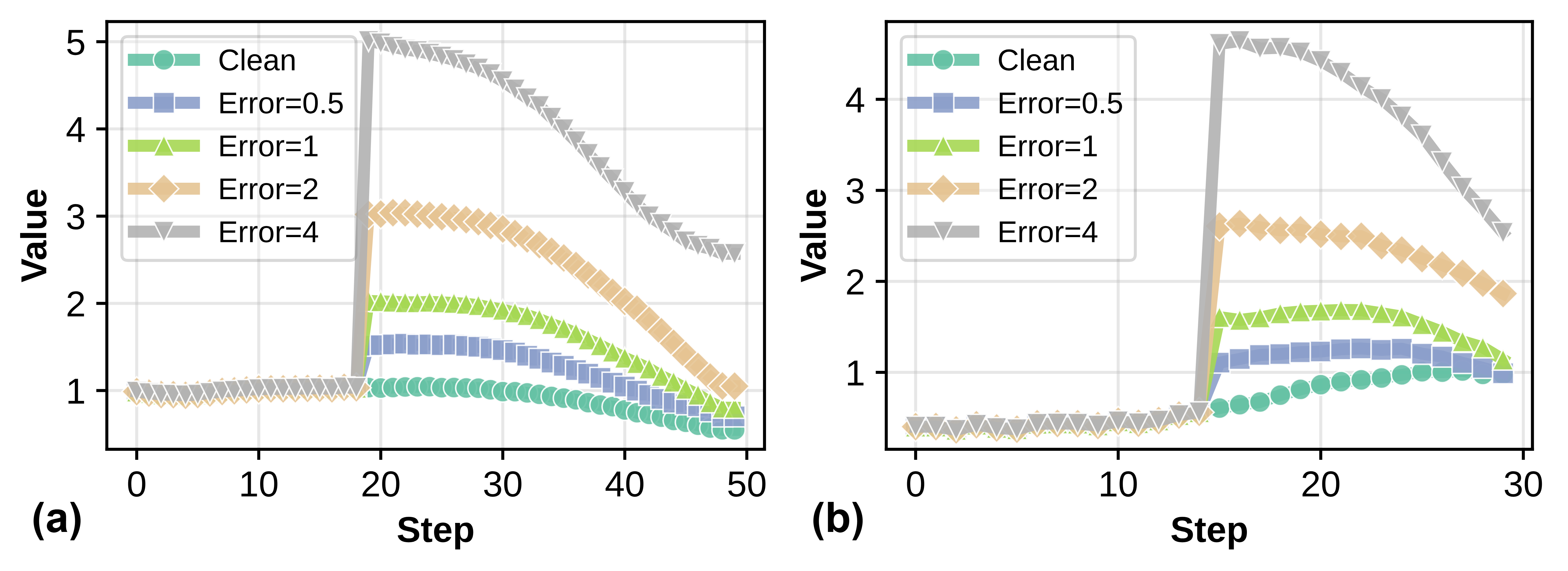}
    \vspace{7pt}
    \caption{
        {Self-correction} ability on (a) DiT and (b) PixArt
    }
    \vspace{-0pt}
    \label{fig: 4resilience_characterization4}
\end{figure}

We further inject errors into different blocks within DiT, including individual Transformer blocks and embedding layers, to compare their fault tolerance. As shown in Fig.~\ref{fig: 4resilience_characterization3}, the first Transformer block exhibits markedly lower robustness, while middle and deeper blocks show more moderate degradation. Besides, the embedding layers also exhibit similar or even worse fault resilience compared to the DiT blocks.
We hypothesize that early layers are more sensitive because they operate directly on raw, unstable input features, making them highly vulnerable to perturbations. As depth increases, representations become more abstract and increasingly dominated by residual pathways, whose activation similarity across layers, as highlighted in \cite{tambe2021edgebert, xu2025specee},  reduces the relative impact of injected faults in newly computed Transformer blocks.
Embedding layers, despite having far fewer parameters than an individual DiT block, also show strong sensitivity. This is likely because their outputs are consumed by every DiT block through cross-attention at every timestep, giving them a wide-ranging, global influence. Therefore, we conclude that in DiTs, \textbf{the early blocks and embedding layers are the most fault-sensitive components}.

\vspace{-2pt}
\subsection{Self-Correction Ability}
\label{fault_resilience_selfcorrect}

To understand error propagation in the multi-step diffusion process, we inject errors at an intermediate denoising step. Fig.~\ref{fig: 4resilience_characterization4} tracks a pixel value over timesteps with different scales of error injected. When errors are introduced, we observe an abrupt deviation in the affected pixel values. 
Results show an initial deviation, followed by a trend where the value begins reverting to its clean trajectory. Smaller errors are often fully corrected at the final step.
These observations indicate that \textbf{diffusion models exhibit inherent self-recovery behavior}: the iterative denoising process tends to identify abnormal deviations and gradually smooth them out. Although artifacts may persist in the final image, the overall process demonstrates a natural capacity for error mitigation, reflecting dynamic robustness against transient errors.


%% file: docs/5_Method.tex
\vspace{-3pt}
\section{DRIFT Framework}
\begin{figure}[t]
    \centering
    \includegraphics[width=1\linewidth]{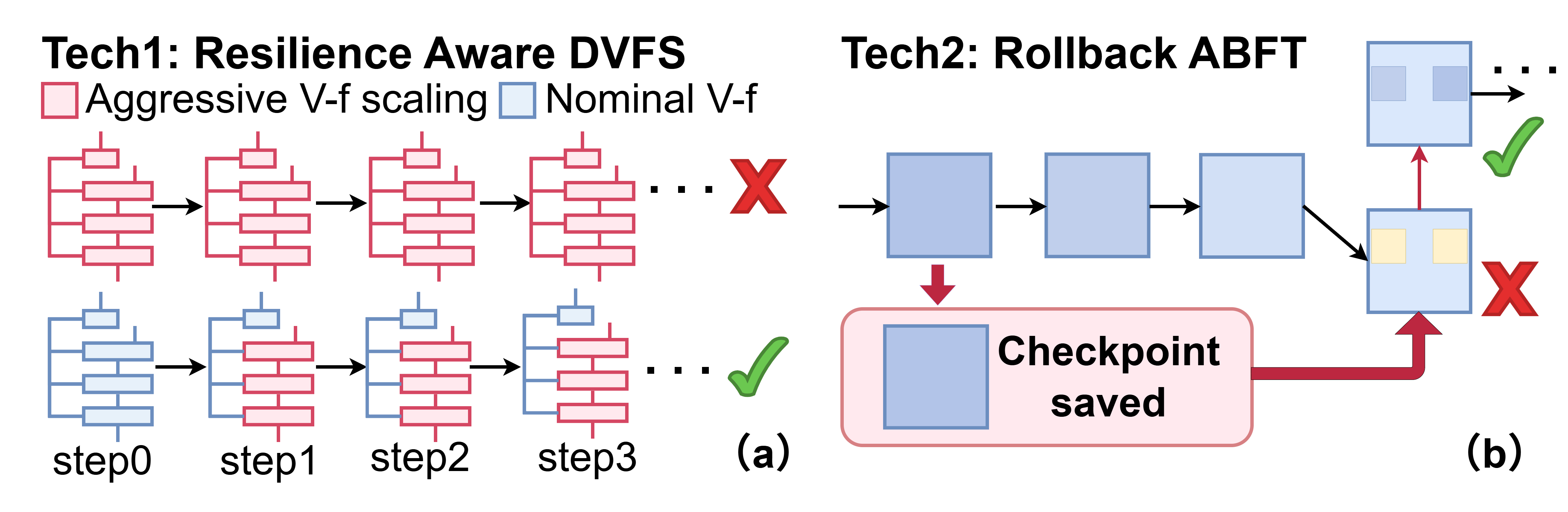}
    \vspace{-0pt}
    \caption{Core techniques in DRIFT. (a) Fine-grained resilience-aware DVFS. (b) Rollback-ABFT mechanism.}
    \vspace{-0pt}
    \label{fig:4.5techs}
\end{figure}

\vspace{-1pt}
\label{sec:methodology}
Building on the above observations, in this section, we introduce our DRIFT framework, which consists of two core techniques, as shown in Fig.~\ref{fig:4.5techs}. 
(i) \textbf{Fine-grained DVFS}, which adaptively applies voltage and frequency across the diffusion process according to the resilience heterogeneity. 
(ii) \textbf{Rollback-ABFT}, which stores lightweight checkpoints and recovers critical errors identified by ABFT.
We further optimize the offloading interval and reorganize the data layout to reduce additional memory overhead.

\vspace{-3pt}
\subsection{Overall Architecture}
\vspace{-1pt}
\label{sec:overall_structure}

\begin{figure}[t]
    \centering
    \includegraphics[width=1.0\linewidth]{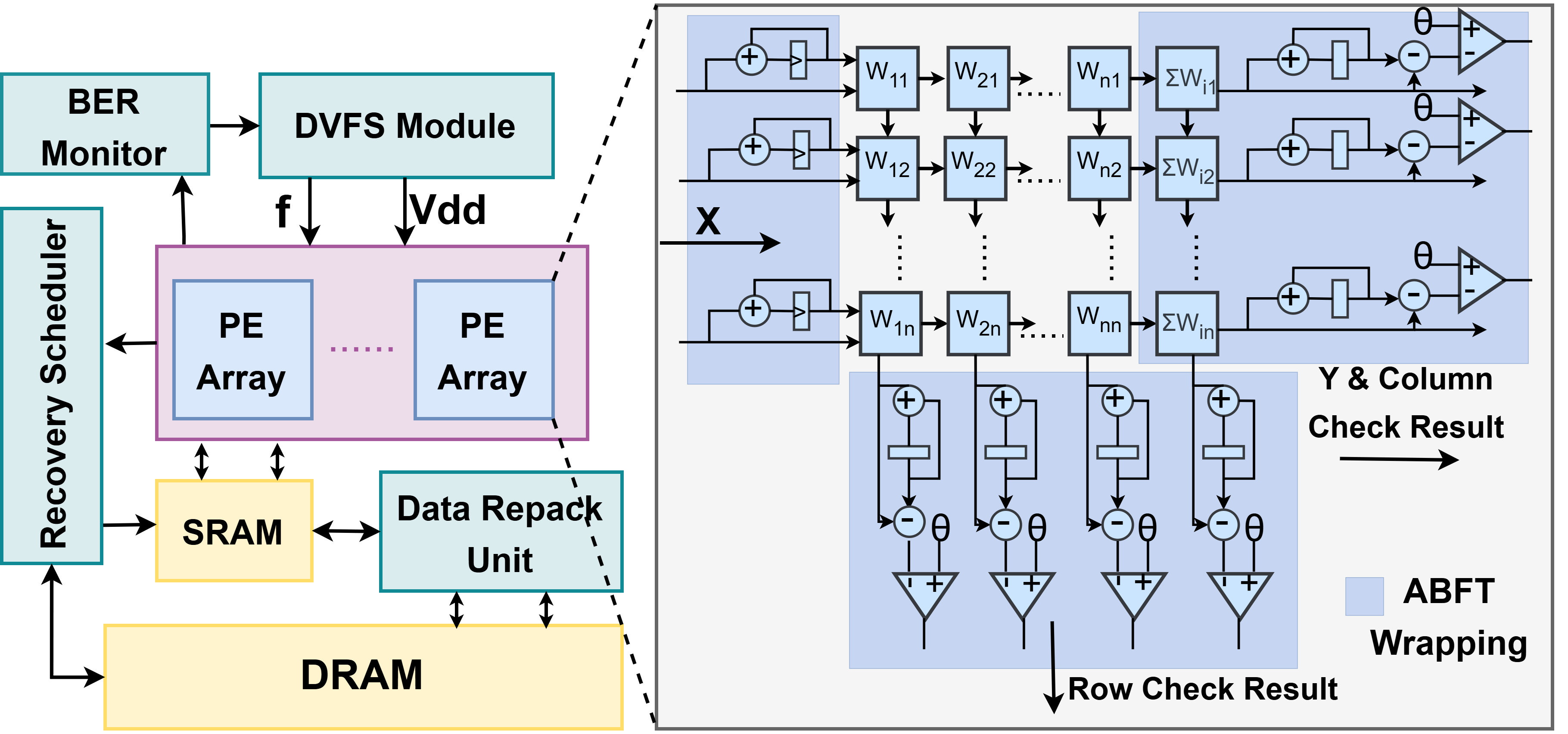}
    \vspace{-0pt}
    \caption{Architecture design for DRIFT. 
    }
    \vspace{-0pt}
    \label{fig:5system_overview}
\end{figure}

As in Fig.~\ref{fig:5system_overview}, our framework is based on a TPU-like accelerator, mainly comprising three components. (i) \textbf{Compute Units}: perform tiled GEMM operations with systolic arrays. (ii) \textbf{On-chip SRAM Buffer}: caches weights and partial results on-demand. (iii) \textbf{Offchip DRAM}: stores full model weights and offloaded activations.
To support our mechanism, we introduce the following components:
\vspace{-1pt}
{\setlength{\leftmargini}{1em}
\begin{itemize}
    \item \textbf{ABFT-Wrapping}: Auxiliary circuits that wrap systolic arrays to compute checksums for ABFT. After GEMM operations, they identify elements with errors exceeding a threshold and report their row and column indices (Sec.~\ref{sec:rollback_abft}).
    \item \textbf{DVFS Module}: Includes an on-chip LDO \cite{kim20210} and an ADPLL \cite{ajayi2020open}, enabling adjustable voltage and frequency.
    \item \textbf{Data Repack Unit}: Reorganizes activations before offloading to DRAM to improve access contiguity.
    \item \textbf{Recovery Scheduler}: Converts error indices into recovery masks, coalesces memory accesses, and manages data dependencies.
    \item \textbf{BER Monitor}: Tracks runtime BERs reported by ABFT and guides the DVFS module to maintain an optimal operating point.
\end{itemize}
}

\vspace{-3pt}
\subsection{Fine-Grained Resilience-Aware DVFS}
\label{sec:fine_grained_dvfs}
\vspace{-1pt}





Aggressive DVFS can improve inference efficiency but risks degrading output quality. We therefore employ a fine-grained, resilience-aware DVFS strategy that adapts operating points to the resilience of different computations (Fig.~\ref{fig:4.5techs}(a)). Based on the analysis in Sec.~\ref{fault_resilience_t} and~\ref{fault_resilience_block}, the early denoising timesteps and the embedding layer are classified as \textit{error-sensitive}, whereas the remaining timesteps and layers are \textit{error-resilient}. Accordingly, the DVFS module assigns \textit{nominal} operating points for error-sensitive computations, which have a small computational footprint but a significant impact on performance. For resilient computations, which constitute the majority of the workload, it applies \textit{aggressive} settings (lower voltage or higher frequency) to leverage their fault tolerance. This strategy significantly enhances efficiency while maintaining acceptable generation quality.

\vspace{-5pt}
\subsection{Error Mitigation via Rollback-ABFT}
\label{sec:rollback_abft}
\vspace{-2pt}

Despite our fine-grained, resilience-aware DVFS, we still observe noticeable quality degradation under higher BERs (e.g., $>10^{-5}$), which limits further efficiency gains. Existing recomputation-based recovery methods \cite{xie2025realm, khoshavi2020shieldenn} fail to leverage the characteristics of diffusion models and incur frequent recomputations at elevated BERs, resulting in suboptimal efficiency.

Leveraging the fault tolerance (Sec.~\ref{fault_resilience_bit} and~\ref{fault_resilience_selfcorrect}) and temporal similarity (Sec.~\ref{bg_diffusion}), we propose a rollback-ABFT mechanism that approximates large-error correction by replacing them with corresponding values from a previous timestep, as depicted in Fig.~\ref{fig:4.5techs}(b). 
We restore only large errors to preserves activation updates and avoid excessive memory access, given that minor errors have a negligible impact (Sec.~\ref{fault_resilience_bit}). The workflow is illustrated as follows:
\vspace{-1pt}
{\setlength{\leftmargini}{1em}
\begin{itemize}
    \item \textbf{Step 1}: We first periodically offload the results of GEMMs to DRAM as checkpoints. This offloading can be overlapped with computation, as diffusion models are compute-bound.
    \item \textbf{Step 2}: During GEMM operations, ABFT detects large computational errors by verifying row and column checksums against a threshold determined according to Sec. \ref{fault_resilience_bit}, (e.g., 10th-bit flips for DiT). Detected error indices are forwarded to the recovery scheduler. While ABFT cannot theoretically distinguish between single large errors and paired large errors that cancel each other within the same row or column, we assume such paired events are statistically negligible under near-random timing errors.
    \item \textbf{Step 3}: The recovery scheduler cross-combines the reported row and column indices to generate a correction mask covering all potential error locations (Fig.~\ref{fig:6method_details}(a)).
    \item \textbf{Step 4}: The scheduler then retrieves the corresponding checkpoint from DRAM and overwrites the masked positions in the current results.
\end{itemize}
}
\vspace{-5pt}
\subsection{Memory Access Optimization}
\vspace{-2pt}
\label{sec:dataflow}

\begin{figure}[!tb]
    \centering
    \includegraphics[width=1.0\linewidth]{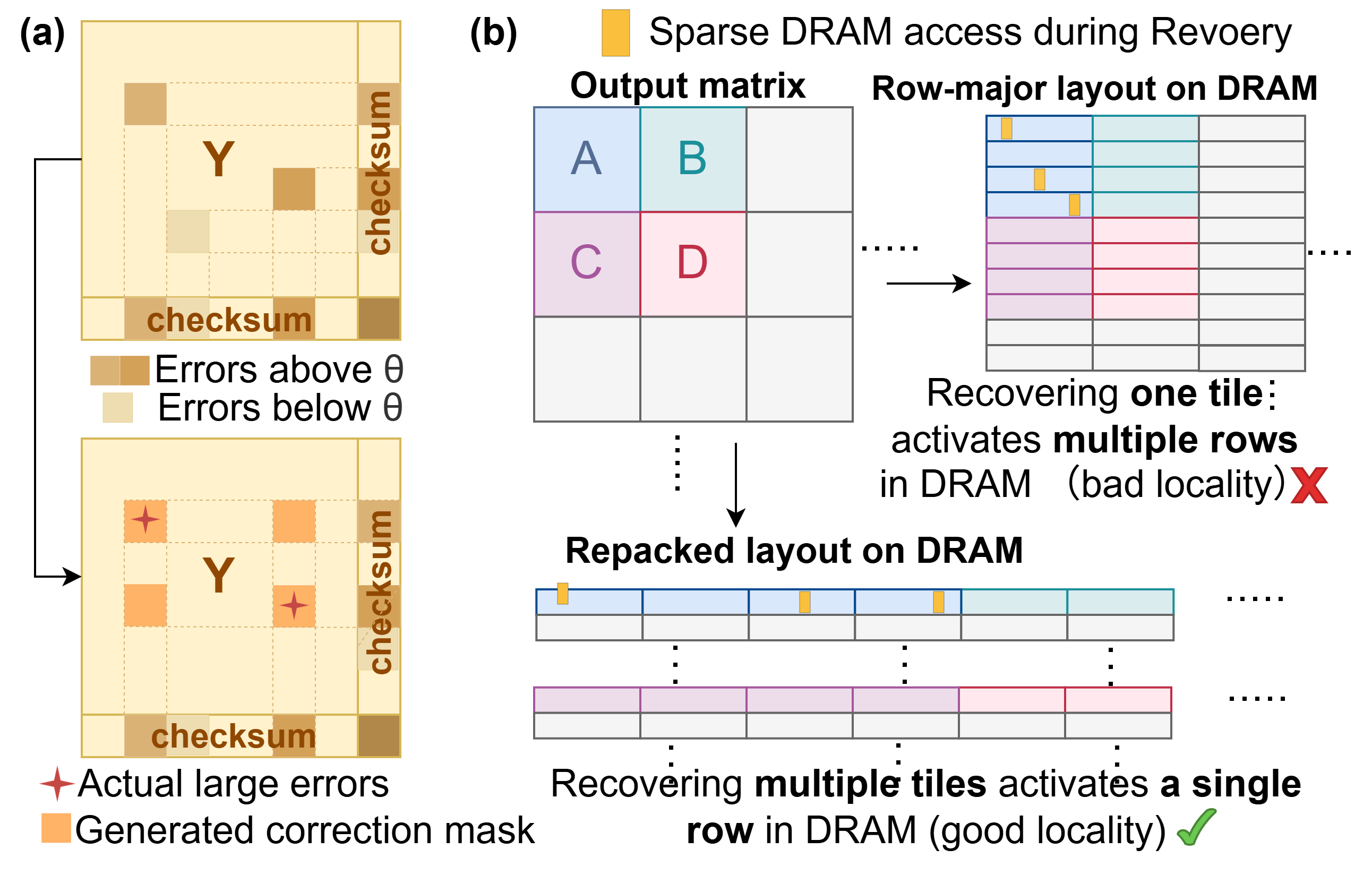}
    \vspace{-0pt}
    \caption{Details in DRIFT techniques.
    (a) Correction mask generation for large errors. (b) Data repacking to reduce redundant DRAM row activations. 
    }
    \vspace{-2pt}
    \label{fig:6method_details}
\end{figure}

While effective at mitigating large errors, a naive implementation of rollback-ABFT faces two key difficulties. (i) Checkpoint offloading overhead: despite diffusion models being compute-bound, their activation sizes are often comparable to weights \cite{kong2024cambricon, qi2025mhdiff}, resulting in substantial DRAM traffic during checkpoint offloading. {(ii) Sparse and fragmented DRAM retrieval}: large-error correction typically requires only a small, scattered subset of activations, whereas DRAM reads operate at row granularity \cite{jun2017hbm}. As in Fig. \ref{fig:6method_details}(b), under conventional data layouts, tile-wise error mitigation triggers excessive and redundant DRAM row activations, causing significant inefficiency. Therefore, we introduce the following optimizations.    




\textbf{Checkpointing at interval}. Given the strong similarity of activations across timesteps, rolling back to a checkpoint from several iterations earlier still provides a fairly accurate approximation for error mitigation. Thus, instead of offloading activations every step, we update the checkpoint only once every $n$ steps, reducing the offloading overhead to $1/n$. We set $n=10$ as discussed in Sec.~\ref{experiment_ablation}.

\textbf{Data layout repacking}. 
As demonstrated in Fig. \ref{fig:6method_details}(b), conventional row-major layouts often scatter elements of the same tile across multiple DRAM rows. Because recovery is performed tile-by-tile, such fragmentation can trigger multiple DRAM-row activations for correcting a single tile. To improve spatial locality, we repack each tile into a 1-D contiguous layout so that, whenever possible, all elements of a tile reside within the same DRAM row. 

\textbf{Memory retrieval overlapping}.
During recovery, memory retrieval can be fully overlapped with computation because adjacent tiles are independent. Recovery only needs to be completed before the next GEMM operation consumes the corresponding tile. 
This requirement is generally satisfied because diffusion models are compute-bound, leaving sufficient bubbles for checkpoint retrieval.

With the above optimizations, the additional memory access overhead remains modest, as detailed in Sec.~\ref{experiment_quality_efficiency}.

\vspace{-10pt}

%% file: docs/6_Evaluation.tex
\section{Experimantal Results}
\label{sec: experiments}

\subsection{Experimental Setup}
\label{experiment_setup}
\paragraph{Hardware Configuration}
\label{experiment_setup_hardware}
We synthesize the systolic arrays (8-bit multipliers and 32-bit accumulators), along with the proposed ABFT-wrapping, data-repack unit, recovery scheduler, and BER monitor using a commercial 14nm PDK. The nominal operating voltage and frequency are 0.9 V and 2 GHz, respectively. 
We assume 64 systolic arrays as compute units and use HBM2 \cite{jun2017hbm} for off-chip memory. Power and area are reported based on synthesis results and \cite{kim20210, ajayi2020open}.
Cycle-level behaviors, including inference latency and memory access, are simulated with SCALE-Sim \cite{samajdar2018scale}.


\paragraph{Models and Datasets.}
We evaluate our method on four diverse configurations.
Specifically, we consider: (1) DiT-XL-512 \cite{peebles2023scalable} (unconditional DiT) on ImageNet \cite{deng2009imagenet}; (2) PixArt-alpha \cite{chen2023pixart} (conditional DiT) on COCO val2017 \cite{lin2014microsoft}; (3) PixArt-alpha on DrawBench \cite{saharia2022photorealistic}; and (4) Stable Diffusion v1.5 \cite{rombach2022high} (conditional UNet) on COCO val2017. We report CLIP, ImageReward, and LPIPS scores to evaluate generation quality comprehensively. In Sec. \ref{experiment_comp_with_previous}-\ref{experiment_taylorseer}, we show results in the first configuration as an example.





\paragraph{DRIFT Configuration.}
Unless otherwise specified, we adopt the following default settings: a systolic array size of 32, an ABFT threshold of $\theta$ corresponding to the 10th bit, and an offloading interval of $n=10$. We assign nominal operating conditions (0.9V, 2GHz) to the timestep embedding and the first 2 denoising steps. For all other computations, we target $BER \approx 3\times10^{-3}$ by using either an undervolted setting (0.68V, 2GHz) to evaluate energy savings or an overclocked setting (0.88V, 3.5GHz) to evaluate speedup,
where 0.88V compensates for increased memory access energy. As shown in Fig.~\ref{fig:6.5results_detailed}(a), our framework also supports a flexible tradeoff between energy efficiency and inference latency.






\vspace{-5pt}
\subsection{Generation Quality and Inference Efficiency}
\label{experiment_quality_efficiency}
\begin{figure}[!tb]
    \centering
    \includegraphics[width=1.0\linewidth]{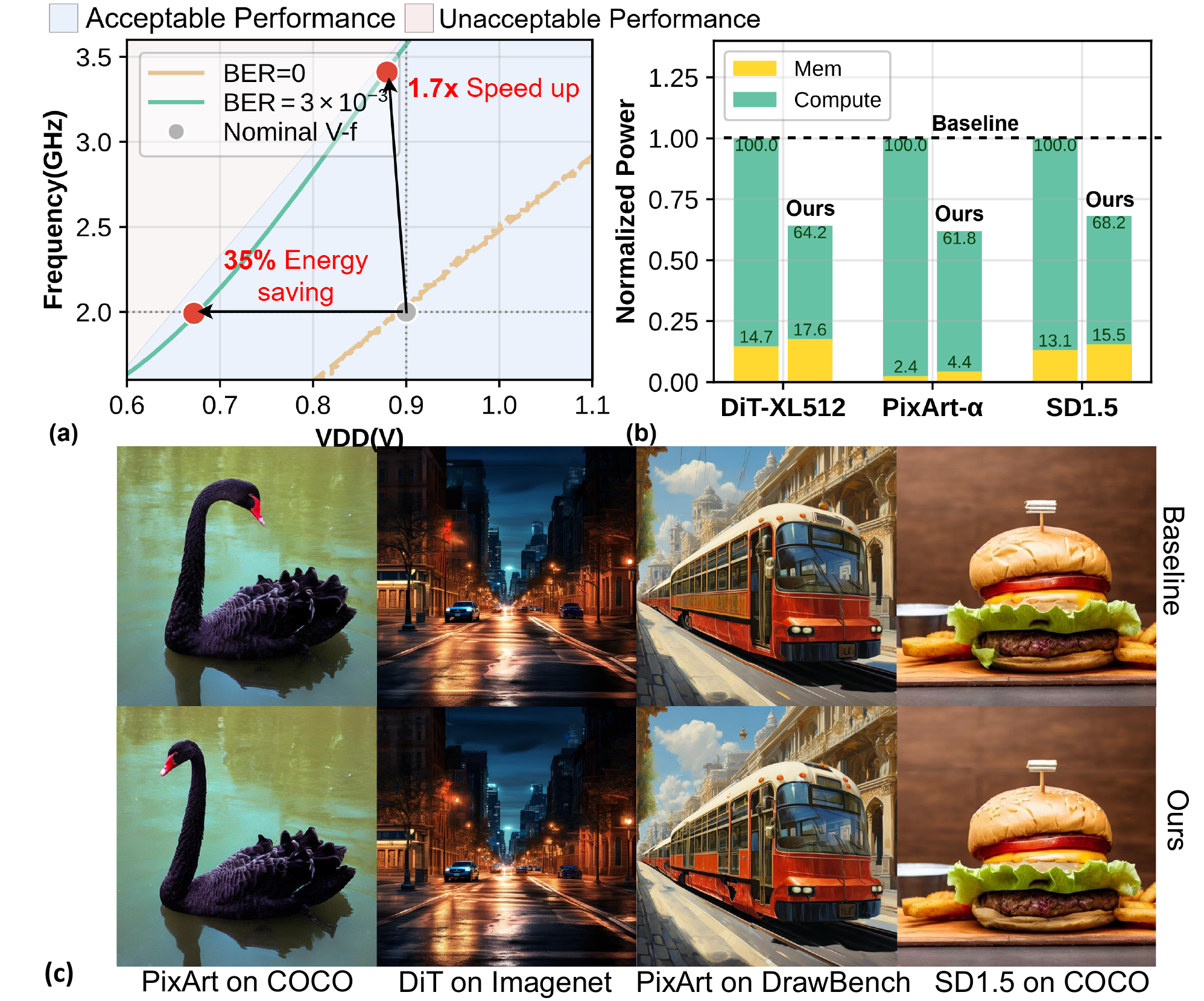}
    \vspace{-0pt}
    \caption{(a) DRIFT can achieve 35\% energy savings via undervolting or 1.7$\times$ speedup via overclocking. (b) Energy breakdown under undervolting. (c) Example generated images.
    }
    \vspace{4pt}
    \label{fig:6.5results_detailed}
\end{figure}

\begin{table}[!tp]
    \centering
    \caption{
    Generation quality and efficiency improvements
    }
    \vspace{4pt}
    \label{tab:main_results}
    \renewcommand*{\arraystretch}{1.1}
    \resizebox{1.0\linewidth}{!}{
    \begin{tabular}{cc|ccc|c|c}  
        \hline\hline
        \textbf{Config}  & \textbf{DRIFT}   & \textbf{CLIP $\uparrow$} & \textbf{ IR$\uparrow$} & \textbf{LPIPS$\downarrow$} & \textbf{Energy/J$\downarrow$} & \textbf{Latency/s$\downarrow$} \\
        \hline\hline
         
         DiT on & w/o   & 0.278 & 0.382 & / & 6.02 & 0.56\\
        ImageNet&  w/    & 0.279 & 0.372 & 0.115 & 3.86(-35.9\%) & 0.33(1.71$\times$)  \\  \hline
         PixArt on   & w/o     & 0.262 & 0.768& / & 28.55 & 2.32 \\
         COCO      & w/     & 0.262& 0.777 & 0.226 & 17.63(-38.3\%) & 1.39(1.67$\times$) \\ \hline
        PixArt on   & w/o    & 0.265 & 0.709 & / & 35.68 & 2.78 \\
         DrawBench   & w/    & 0.264 & 0.714 & 0.193 & 22.05(-38.2\%) & 1.62(1.70$\times$) \\ \hline
         SD1.5 on& w/o   & 0.265& 0.189 & / & 2.71 & 0.77  \\
         COCO & w/    & 0.267& 0.187 & 0.155 & 1.86(-31.2\%)
         & 0.47(1.66$\times$) \\   \hline

        \hline\hline
    \end{tabular}
    }
     \vspace{-10pt}
\end{table}


The ABFT wrapper incurs approximately 6.3\% additional power and energy, while other components in Sec. \ref {sec:overall_structure} are negligible. 
Table~\ref{tab:main_results} reports the image quality and efficiency improvements achieved by DRIFT (with FID\cite{heusel2017gans} on DiT-XL512: baseline 3.578; DRIFT 3.586). Compared with nominal operations, DRIFT maintains nearly identical CLIP and ImageReward scores, indicating a negligible impact on semantic fidelity and human preference. LPIPS remains comparable to existing efficient algorithms \cite{TaylorSeer2025}, reflecting only minor pixel-level differences. While preserving generation quality, DRIFT enables 36\% energy saving via undervolting and 1.7$\times$ speedup by overclocking on average.
Fig.~\ref{fig:6.5results_detailed}(c) shows image examples of the baseline and DRIFT.


\vspace{-3pt}
\paragraph{\textbf{Memory Overhead Analysis.}}
Fig.~\ref{fig:6.5results_detailed}(b) shows the energy breakdown for the undervolted setting. 
DRIFT incurs two types of off-chip memory overhead: extra DRAM writes for checkpointing every $n$ steps and extra DRAM reads for recovery at each step. We estimate the recovery cost from the required DRAM row activations and cache-line reads. For DiT-XL-512 under the Sec. \ref{experiment_setup} configuration, each contributes about 10\% additional memory access. Since diffusion-model inference is compute-bound, however, the overall energy overhead remains below 3\%.




\begin{figure}[!tb]
    \centering
    \includegraphics[width=0.9\linewidth]{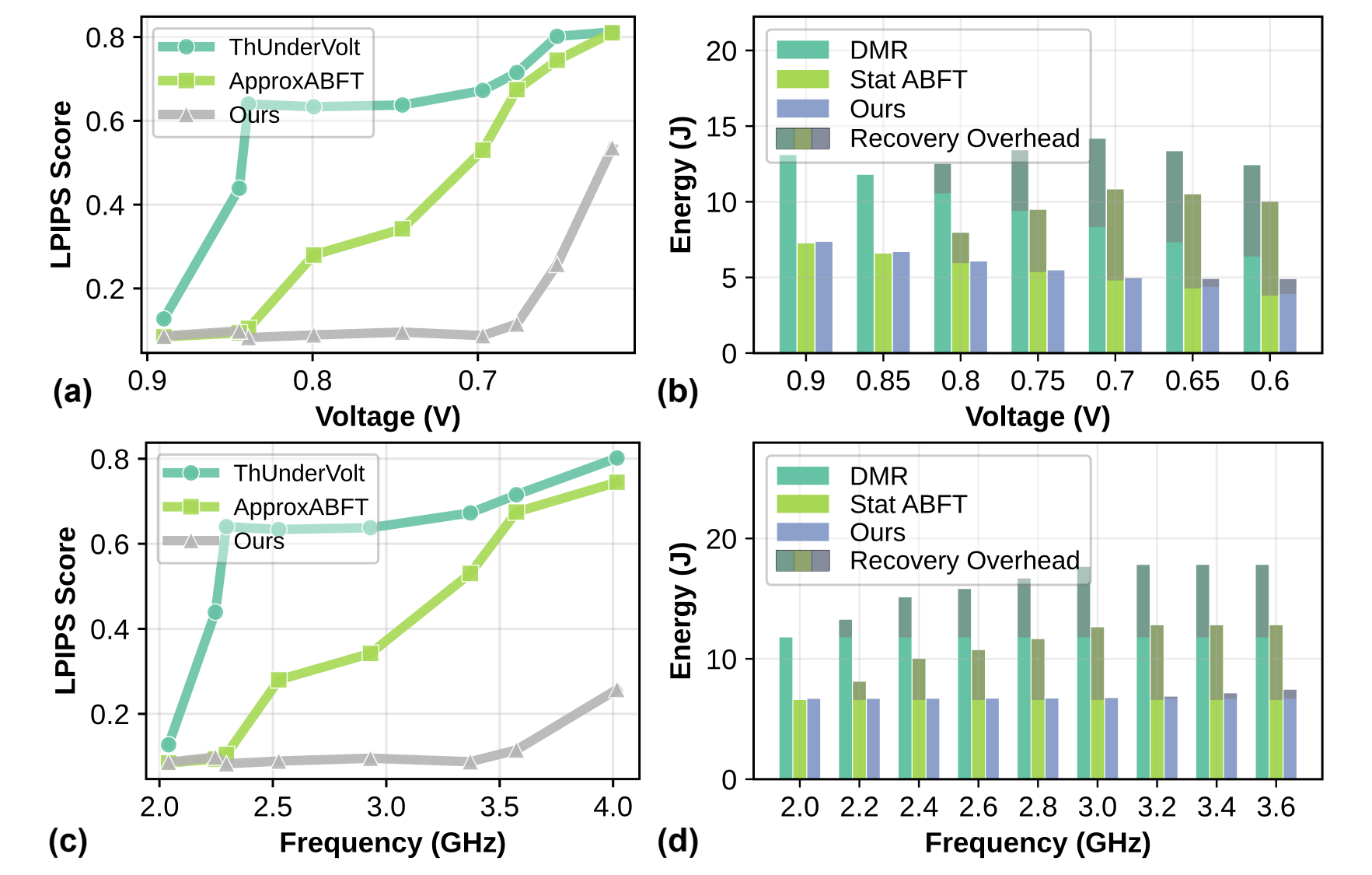}
    \vspace{8pt}
    \caption{Comparison with previous works. (a)(c) DRIFT outperforms ThUnderVolt \cite{zhang2018thundervolt} and ApproxABFT \cite{xue2023approxabft} in reliability enhancement, and (b)(d) surpasses DMR and Stat ABFT \cite{xie2025realm} in recovery efficiency.
    }
    \vspace{-0pt}
    \label{fig:7previous works}
\end{figure}

\begin{figure}[!tb]
    \centering
    \includegraphics[width=0.9\linewidth]{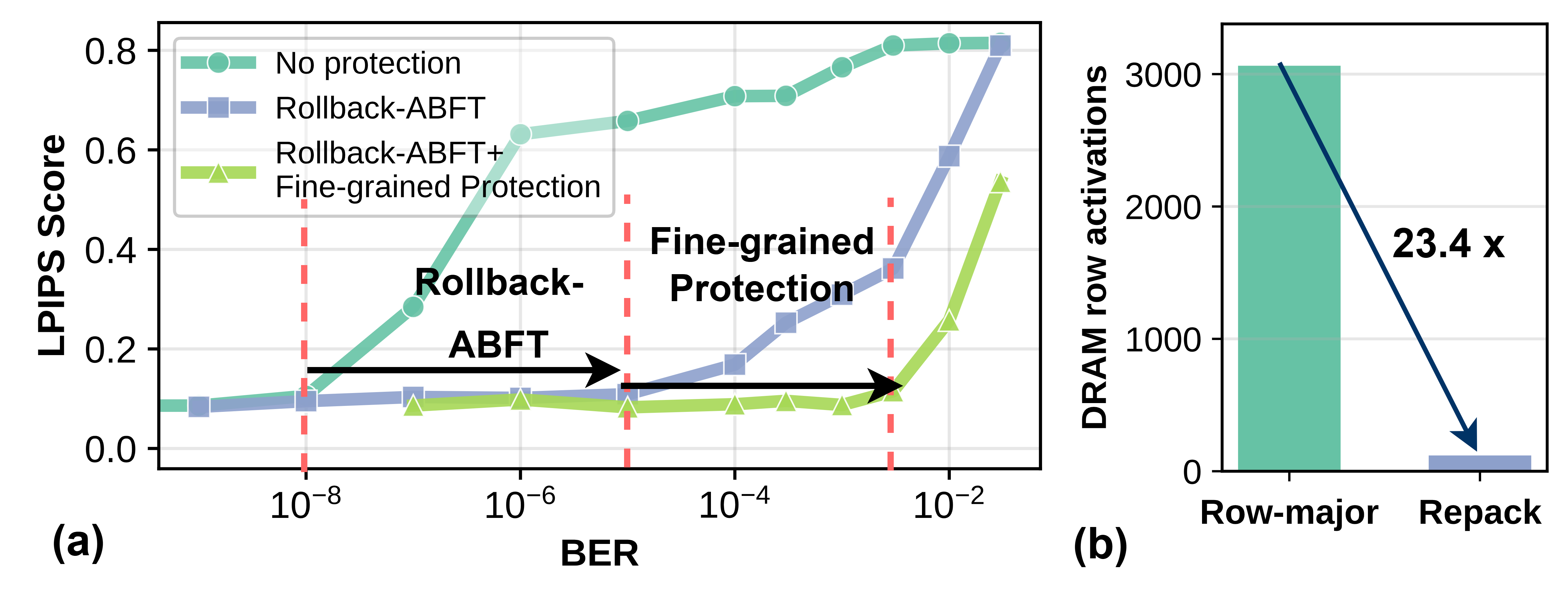}
    \vspace{5pt}
    \caption{Evaluation of (a) fine-grained resilience-aware DVFS, rollback-ABFT, and (b) data layout repacking.
    }
    \vspace{-0pt}
    \label{fig:8ablation}
\end{figure}

\begin{figure}[!tb]
    \centering
    \includegraphics[width=0.9\linewidth]{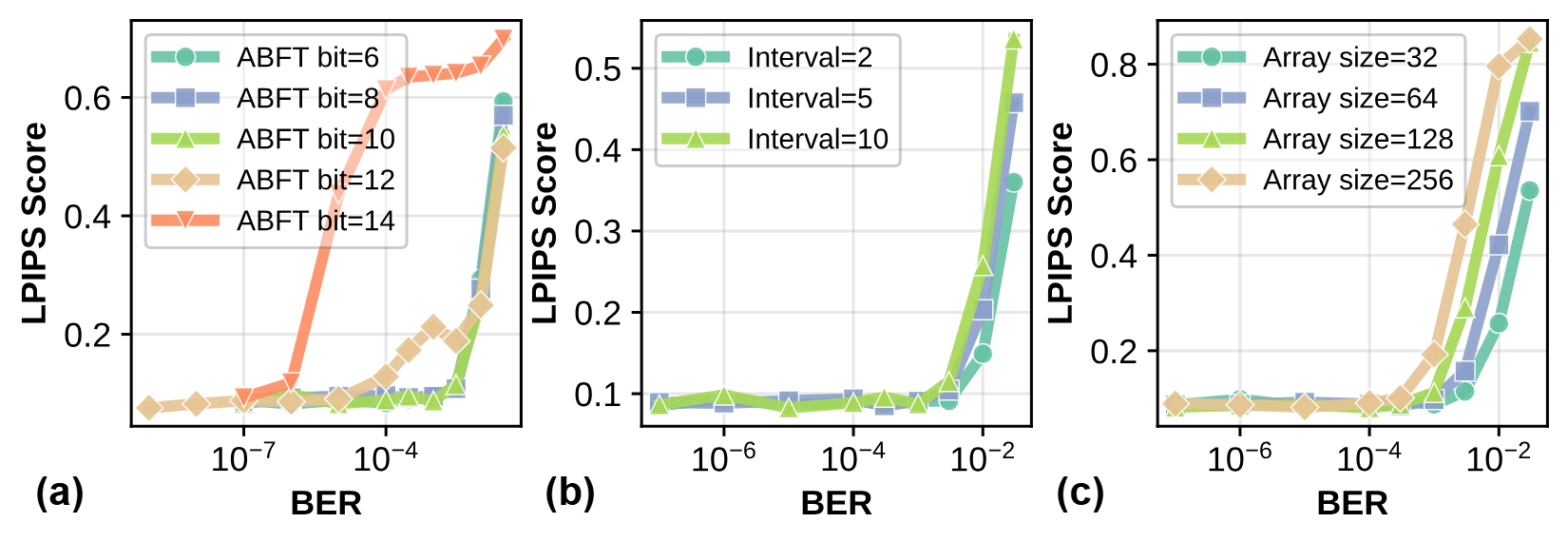}
    \vspace{8pt}
    \caption{Design space exploration on (a) ABFT threshold, (b) offloading interval, and  (c) systolic array size.
    }
    \vspace{-0pt}
    \label{fig:9dse}
\end{figure}

\vspace{-6pt}
\subsection{Comparison with Previous Works}
\label{experiment_comp_with_previous}
This section compares DRIFT with existing error mitigation techniques for undervolting (Fig.~\ref{fig:7previous works}(a)(b)) and overclocking (Fig.~\ref{fig:7previous works}(c)(d)). DRIFT consistently outperforms prior approaches in both reliability enhancement and recovery efficiency. For reliability enhancement (Fig.~\ref{fig:7previous works}(a)(c)), ThUnderVolt \cite{zhang2018thundervolt} skips faulty computations and ApproxABFT \cite{xue2023approxabft} zeros out detected anomalies. Both approaches fail to maintain generation quality at higher BERs due to excessive neuron pruning. For recovery efficiency (Fig.~\ref{fig:7previous works}(b)(d)), DMR and stat ABFT \cite{xie2025realm} trigger recomputation to ensure correctness. Although effective at preserving image quality, they incur frequent recoveries and therefore substantial overhead. 

\vspace{-2pt}
\subsection{Ablation Study}
\label{experiment_ablation}
This subsection evaluates individual techniques in our framework. As shown in Fig.~\ref{fig:8ablation}(a), the rollback-ABFT mechanism relaxes the BER tolerance for comparable image quality from $1\times10^{-8}$ to $1\times10^{-5}$, while fine-grained protection further extends to $3\times10^{-3}$. This relaxation enables the aggressive DVFS discussed above. Fig.~\ref{fig:8ablation}(b) illustrates the effectiveness of our data-layout repacking: for q\_proj in DiT, it reduces DRAM row activations by 23.4$\times$. Consequently, computation takes approximately 15$\upmu \rm s$, while memory retrieval incurs only 714ns, and thus can be fully overlapped.





\vspace{-2pt}
\subsection{Design Space Exploration}
\label{experiment_dse}

This subsection examines the impact of parameters involved in DRIFT with DiT. As shown in Fig.~\ref{fig:9dse}(a), ABFT thresholds above the 10th bit fail to preserve output quality because critical errors escape detection, aligning with Sec.~\ref{fault_resilience_bit}. In Fig.~\ref{fig:9dse}(b), offloading intervals of 2, 5, and 10 yield comparable performance, whereas shorter intervals incur higher offloading overhead. Our techniques remain consistently effective across different systolic-array sizes (Fig.~\ref{fig:9dse}(c)), demonstrating broad applicability across diverse hardware configurations.



\vspace{-2pt}
\subsection{Compatibility with Efficient Algorithms}
\label{experiment_taylorseer}

To evaluate compatibility with efficient algorithms, we integrate DRIFT with a recent cache-based algorithm, TaylorSeer \cite{TaylorSeer2025}, using an interval of 3 and a cache order of 2. Table~\ref{tab:10x10_table} reveals that, with 2 nominal steps and 16 overclocked steps, their combination achieves a 4.40$\times$ speedup while maintaining generation quality. These results indicate that DRIFT is orthogonal to existing efficient algorithms and can be seamlessly combined with them for further benefits.

\vspace{-9pt}
\begin{table}[htp]
    \centering
    \caption{Combination with TaylorSeer}
    \vspace{4pt}
    \label{tab:10x10_table}
    \renewcommand*{\arraystretch}{1.05}
    \resizebox{1.0\linewidth}{!}{
    \begin{tabular}{c|ccc|c}  
        \hline\hline
        \textbf{Methods}    & \textbf{CLIP $\uparrow$} & \textbf{ ImageReward$\uparrow$} & \textbf{LPIPS$\downarrow$} &  \textbf{Speedup$\uparrow$} \\
        \hline\hline
        Baseline&0.2789&0.3715& / & 1.00$\times$\\
        TaylorSeer&0.2791&0.3897&0.1708&2.82$\times$ \\
        DRIFT&0.2793&0.3715&0.1149& 1.71$\times$\\
        TaylorSeer + DRIRT&0.2785&0.3793&0.1827&4.40$\times$ \\

        \hline\hline
    \end{tabular}
    }
\end{table}
\vspace{-15pt}

%% file: docs/7_Conclusions.tex
\section{Conclusion}

We propose DRIFT, an algorithm–architecture co-design framework that exploits the inherent fault tolerance of diffusion models to enable aggressive DVFS for efficient and reliable inference. We conduct extensive error-injection studies to characterize the resilience of the diffusion generation process,
DRIFT incorporates (1) fine-grained, resilience-aware DVFS to protect vulnerable network blocks, and (2) a rollback-ABFT mechanism that filters and selectively corrects large errors by reverting to previous latent states. We further optimize memory overhead through extended offloading intervals and data layout repacking. DRIFT achieves 36\% average energy savings via undervolting and a 1.7$\times$ speedup via overclocking without compromising output quality. As a hardware-oriented method, DRIFT is orthogonal to existing efficient algorithms, offering a universal path to enhance diffusion model deployment.